\documentclass[aps,prd,reprint,preprintnumbers,floats,epsfig,nofootinbib,amssymb,
nofootinbib]{revtex4}

\usepackage{slashed}
\usepackage{graphicx,color}
\usepackage{epsfig}
\usepackage{subfigure}
\usepackage{epsfig}
\usepackage{epstopdf}
\usepackage{dcolumn}
\usepackage{bm}
\usepackage{color}
\usepackage{ulem}
\usepackage{enumitem}
\usepackage[colorlinks,
            linkcolor=black,
            anchorcolor=black,
            citecolor=black
            ]{hyperref}

\begin{document}

\baselineskip=15pt


\title{Flavor Specific $U(1)_{B_q - L_\mu}$ Gauge Model for Muon $g-2$ and $b \to s \bar \mu \mu$ Anomalies}

\author{Jian-Yong Cen$^{1}$\footnote{cenjianyong@163.com},
Yu Cheng$^{2,3}$\footnote{chengyu@sjtu.edu.cn},
Xiao-Gang He$^{2,3,4}$\footnote{hexg@phys.ntu.edu.tw},
Jin Sun$^{2,3}$\footnote{019072910096@sjtu.edu.cn}}

\affiliation{${}^{1}$School of Physics and Information Engineering,
Shanxi Normal University, Linfen, Shanxi 041004, China}
\affiliation{${}^{2}$Tsung-Dao Lee Institute \& School of Physics and Astronomy, Shanghai Jiao Tong University, Shanghai 200240, China}
\affiliation{${}^{3}$Shanghai Key Laboratory for Particle Physics and Cosmology, Key Laboratory for Particle Astrophysics and Cosmology (MOE), Shanghai Jiao Tong University, Shanghai 200240, China}
\affiliation{${}^{4}$Department of Physics, National Taiwan University, Taipei 10617, Taiwan}

\begin{abstract}
The muon $(g-2)_\mu$ and $b\to s \bar \mu \mu$ induced $B$ anomalies as hints of new physics beyond the standard model (SM) have attracted much attention. These two anomalies indicate that there may exist new interaction specifically related to muon. A lot of theoretical ideas have been proposed to explain these anomalies. Gauged flavor specific $U(1)_{B_q-L_\mu}$ is among the promising ones. The new gauge boson $Z'$ from $U(1)_{B_q-L_\mu}$ interacts with muon and provides necessary ingredient to solve the $(g-2)_\mu$ anomaly. The $Z'$-quark coupling can generate flavor changing interactions after diagonalization of quark mass matrix between weak eigen-state and mass eigen-state basis.
We revisit challenges for such models attempting to explain the $(g-2)_\mu$ and $B$ anomalies separately or simultaneously. We find although for $U(1)_{B_q-L_\mu}$ models there is still parameter space to provide solutions for
separately explaining the $(g-2)_\mu$ and $B$ anomalies, there exists no parameter space for such models to solve both the anomalies simultaneously, after taking into account existing constraints from $\tau \to \mu \gamma$, $\tau \to 3 \mu$, neutrino trident and $B_s - \bar B_s$ data.   Among them leptonic processes restrict $Z^\prime$ mass to be less than a few hundred MeV if required to solve the $(g-2)_\mu$ anomaly, which causes conflict between data from  $B_s - \bar B_s$, $D^0 - \bar D^0$ mixing and also hadron decays with $Z^\prime$ in the final states.
The effects of $U(1)_Y$ and $U(1)_{B_q-L_\mu}$ kinetic mixing  on these anomalies are also studied. We find that neither can these effects  do much to bring the two anomalies together to be solved simultaneously.
\end{abstract}

\maketitle

\section{Introduction}


The Muon $g-2$ Collaboration at Fermilab reported their new results from Run 1 measurement of the muon
anomalous magnetic dipole moment $a_\mu$ recently~\cite{Fermilab}. Combining previous data from BNL~\cite{previous-data}, the discrepancy between
experimental data $a^{exp}_\mu$ and SM  prediction $a^{SM}_\mu$~\cite{SM-number} reinforces $(g-2)_\mu$ of muon anomaly confidence level which raised from 3.7$\sigma$ to 4.2$\sigma$~\cite{Fermilab} with,
$\Delta a_\mu = a^{exp}_\mu - a^{SM}_\mu = (251\pm 59)\times 10^{-11}$.
Needless to say that more precise SM calculation is needed to further confirm this anomaly\footnote{Recent lattice calculation in fact favors experimental value compared with previous calculations~\cite{nature-g-2}.}. This anomaly generates a new wave of extensive theoretical studies,
including new $U(1)$ gauge models~\cite{gauge1,gauge2,gauge3}, multi-Higgs models~\cite{multi-Higgs1, multi-Higgs2,multi-Higgs3, multi-Higgs4, multi-Higgs5, multi-Higgs6}, axion or axion-like models~\cite{axion1, axion2,axion3,axion4,axion5}, supersymmetric models~\cite{susy00, susy0, susy1,susy2, susy3,susy4,susy5,susy6, susy7, susy8, susy9,susy10,susy11,susy12,susy13, susy14, susy15, susy16, susy17} and many other interesting models~\cite{other00, other0, other1,other2,other3,other4,other5,other6,other7,other8,other9}.

There exist also persistent $B$ anomalies from $b\to s \bar \mu \mu$ induced rare $B$ decay processes, such as the branching ratios of $B\to K \bar \mu \mu,K^* \bar \mu \mu$, $B_s \to \phi \bar \mu \mu, \bar \mu \mu$, and $B\to K^* \bar \mu \mu$ angular distribution, between SM predictions and experimental measurements~\cite{review-fit,review-fit-earlier1,review-fit-earlier2}. The recent full run 2 data from LHCb~\cite{LHCb1} for $R_K = 0.846^{+0.042+0.013}_{-0.039-0.012}$ for the di-muon invariant mass squared $q^2$ between 1.1 GeV$^2$  to 6 GeV$^2$, increased the $R_K$ deviation to 3.1$\sigma$ level.
These anomalies if confirmed also indicate new physics beyond SM and attract  a lot of theoretical attentions~\cite{gauge1, multi-Higgs2, multi-Higgs5, other0, other1, hiller, b-anomaly1, b-anomaly2, b-anomaly3, b-anomaly4, b-anomaly5, b-anomaly6}.

We wonder whether these two cases of anomalies could be correlated with each other from model perspective.
Thus, we need to analyze their particular and common features. Firstly, note that the $(g-2)_\mu$ and $b\to s \bar \mu \mu$ induced $B$ anomalies all involve muon pairs,
which indicates potentially that new physics interactions may be related to the second generation of charged lepton.
Then, the $B$ anomalies may require mixing between the second and third generations of quarks for new physics.
There are different ways to realize such new physics interactions. Exchanging a new gauge boson $Z^\prime$, which results in models beyond SM with an additional  $U(1)$ gauge symmetry, is one of the favored mechanisms.


To have a consistent model with new gauge interactions, one must make sure that the models constructed are gauge anomaly free.
Many models~\cite{Li-Lj1,Li-Lj2, BDHK,2007mixing, biswas, gauge2, gauge3} which can provide solution to $(g-2)_\mu$ anomaly through exchange of a $Z'$ have been proposed. The $U(1)_{L_\mu - L_\tau}$ model is the simplest one of this type. However, it does not involve $Z'$-quark interactions and needs to be extended further to address $B$ anomalies.
Several variations of such models have been proposed to explain the recent muon $g-2$ anomaly~\cite{gauge2, gauge3}.
If right-handed neutrinos are introduced, the gauged $U(1)_{B-L}$ model is also  anomaly free~\cite{gauge-B-L} which can be used as a consistent model to work with. Again in the simplest model, neither is there quark mixing.
In Ref.~\cite{pospelov} some vector-like up and down type of quarks in $U(1)_{L_\mu - L_\tau}$ model were introduced to achieve the goal.
In fact without introducing new type of quarks, flavor changing interactions can also be generated if one assigns non-trivial $U(1)_{L_\mu - L_\tau}$ or $U(1)_{B - L}$ quantum numbers for Higgs bosons which generate masses for quarks and leptons.
When working in the mass eigen-basis, $Z'$ will in general have flavor changing interactions. There are several studies on related models~\cite{crivellin, Alonso:2017uky, ben, greljo}.
For $U(1)_{B-L}$ if all generations have uniform new gauge charges, no flavor changing $Z'$ interaction can be generated. But one or two generations have non-trivial $U(1)_{B-L}$ charges and the other generation has trivial charges, flavor changing $Z'$ interaction with quarks and leptons can be generated in the mass eigen-basis.
Therefore such models may be able to explain $g-2$ anomaly or $b\to s \bar \mu \mu$ induced $B$ anomalies. We refer this as a flavor specific $U(1)_{B_q-L_\mu}$ gauge model. $q$ indicates the quark generation number.
In addition, the kinetic mixing between $U(1)_Y$ of the SM and the new $U(1)$ unavoidably arises, which can also introduce flavor changing $Z'$-fermions interactions.

Gauged flavor specific $U(1)_{B_q-L_\mu}$  model to explain the $(g-2)_\mu$ or the $B$ anomalies has been proposed in the literature \cite{Bian:2017rpg,Bian:2017xzg,Allanach:2020kss,Alonso:2017uky,greljo,Bonilla:2017lsq}.
The new gauge boson $Z'$ interactions with muon and flavor changing quark  can provide the solution for the $(g-2)_\mu$ anomaly and $B$ anomalies, respectively.
It is tempting to see whether such models can explain both anomalies simultaneously. We revisit challenges  for such models  attempting to explain the $(g-2)_\mu$ and $B$ anomalies separately or simultaneously. We find that although for $U(1)_{B_q-L_\mu}$ models there is still parameter space to explain separately  the $(g-2)_\mu$ or $B$ anomalies, there exists no region for such models to solve both the anomalies simultaneously, after taking into account existing constraints from  $\tau \to \mu \gamma$, $\tau \to 3 \mu$, neutrino trident and $B_s - \bar B_s$,  $D^0 - \bar D^0$ mixing and also hadron decays with $Z^\prime$ in the final states, such as  $D \to \pi Z^\prime$.

This paper is organized as follows. In Sec. II, we show the details of flavor specific $U(1)_{B_2-L_\mu}$ model. Sec. III provides the solution to $(g-2)_\mu$ while satisfying other constraints from leptonic processes. Sec. IV explains the $B$ anomalies and other constraints from quark sector.  Sec. V shows the difficulties to explain both anomalies simultaneously for variant $U(1)_{B_q-L_l}$ models. Sec. VI is devoted to studying the effects of $U(1)_Y$ and $U(1)_{B_2-L_\mu}$ kinetic mixing. In Sec. VII, we  draw our conclusion.
\\

\section{The flavor specific $U(1)_{B_2-L_\mu}$ model}

The gauge group of the model is $SU(3)_C\times SU(2)_L\times U(1)_Y\times U(1)_{B_2 -L_\mu}$ with the corresponding coupling constants $g_s$, $g$, $g'$ and $\tilde g$, respectively.
The left-handed quarks, $Q_L = (u_L, d_L)^T = (Q_{L_1},\;Q_{L_2}, \;Q_{L_3})$, the right-handed up type of quarks, $U_R = (u_{R_1},\;u_{R_2}, \;u_{R_3})$, the right-handed down type of quarks, $D_R = (d_{R_1},\;d_{R_2}, \;d_{R_3})$,
the left-handed leptons $L_L = (\nu_L,\;l_L)^T = (L_{L_1},\;L_{L_2}, \;L_{L_3})$,  the right-handed charged leptons, $l_R = (l_{R_1},\;l_{R_2}, \;l_{R_3})$, and the right-handed neutrinos,
$\nu_R = (\nu_{R_1},\;\nu_{R_2}, \;\nu_{R_3})$, have $SU(3)_C\times SU(2)_L\times U(1)_Y$ quantum numbers $(3,2,1/6)$, $(3,1, 2/3)$, $(3,1,-1/3)$, $(1,2,-1/2)$,  $(1,1,-1)$ and $(1,1,0)$, respectively.
Here the subscripts ${1,2,3}$ correspond to  different generations.
Note that the first and third generations do not transform under $U(1)_{B_2-L_\mu}$. Only the second generation of quarks and leptons have $U(1)_{B_2-L_\mu}$ charge with $1/3$ and $-1$, respectively.
The Higgs boson $H$ transforms as $(1,2,1/2)$ and $(0)$ under the SM  and $U(1)_{B_2-L_\mu}$ gauge group.
 The corresponding $U(1)_{B_2-L_\mu}$ charges for all particles in our model are collected in Table I.
\begin{table}
	\caption{  The $U(1)_{B_2-L_\mu}$ charges for all particles.}
	\begin{tabular}{|c|c|c|c|c|c|c|c|c|c|}
		\hline
		Quarks  & $Q_{L_1}$ & $Q_{L_2}$ & $Q_{L_3}$ & $u_{R_1}$  & $u_{R_2}$ & $u_{R_3}$ & $d_{R_1}$  & $d_{R_2}$ & $d_{R_3}$ \\
		\hline
		$U(1)_{B_2-L_\mu}$  & 0  & 1/3  & 0 & 0 & 1/3 & 0 & 0 & 1/3  & 0   \\
		\hline
		Leptons   & $L_{L_1}$ & $L_{L_2}$ & $L_{L_3}$ & $l_{R_1}$ & $l_{R_2}$ & $l_{R_3}$ & $\nu_{R_1}$ & $\nu_{R_2}$ & $\nu_{R_3}$  \\
		\hline
		$U(1)_{B_2-L_\mu}$  & 0  & -1  & 0 & 0 & -1 & 0 & 0 & -1  & 0    \\
		\hline
		Scalars 	& $H$ & $H_1^q$ & $H_2^q$ & $H_1^l$ & $H_2^l$ & $S_1$ & $S_2$ & & \\
		\hline
		$U(1)_{B_2-L_\mu}$    & 0 & 1/3 & -1/3 & 1 & -1 & 1 & 2 & &\\
		\hline
	\end{tabular}
\end{table}

For the above assignments of quantum numbers, the interactions of $Z'$ from  $U(1)_{B_2-L_\mu}$ with fermions are given by
\begin{eqnarray}
L_{int} &=& {1\over 3} \tilde g \left ( \bar Q_{L_2} \gamma^\mu Q_{L_2} +  \bar u_{R_2} \gamma^\mu u_{R_2} +  \bar d_{R_2} \gamma^\mu d_{R_2} \right )Z'_\mu- \tilde g \left ( \bar L_{L_2} \gamma^\mu L_{L_2} +  \bar l_{R_2} \gamma^\mu l_{R_2} +  \bar \nu_{R_2} \gamma^\mu \nu_{R_2} \right )Z'_\mu\;.
\end{eqnarray}
The Yukawa interactions and also the right-handed neutrino mass terms are given by
\begin{eqnarray}
L_{Y-mass} = -\left (\bar Q_L Y^u_H U_R \tilde H + \bar Q_L Y^d_H D_R H + \bar L_L Y^\nu_H \nu_R  \tilde H + \bar L_L Y^l_H l_R  H + {1\over 2} \bar \nu_R^c \tilde M_R \nu_R \right )
+ \mbox{H.c.}\;,
\end{eqnarray}
where $\nu^c_R$ is the charge conjugated field of $\nu_R$. The forms of Yukawa matrix $Y^f_H$ and mass matrix $\tilde M_R$   are
\begin{eqnarray}
Y_H^f = \left ( \begin{array}{ccc}
\;\;Y^f_{11}\;\;&\;\;0\;\;&\;\;Y^f_{13}\;\;\\
\;\;0\;\;&\;\;Y^f_{22}\;\;&\;\;0\;\;\\
\;\;Y^f_{31}\;\;&\;\;0\;\;&\;\;Y^f_{33}\;\;
\end{array}
\right )\;,\;\;\;\;\tilde M_R = \left ( \begin{array}{ccc}
\;\;M_{11}\;\;&\;\;0\;\;&\;\;M_{13}\;\;\\
\;\;0\;\;&\;\;0\;\;&\;\;0\;\;\\
\;\;M_{13}\;\;&\;\;0\;\;&\;\;M_{33}\;\;
\end{array}
\right )\;.
\end{eqnarray}

After $H$ develops vacuum expectation value (VEV) $v_0/\sqrt{2}$, the fermion mass matrices are in the form $M_f = Y^f_H v_0/\sqrt{2}$. The mass matrices with the current form do not produce correct Kobayashi-Maskawa (KM) matrix for quarks and Pontecorvo-Maki-Nakagawa-Sakata (PMNS) mixing matrix for leptons. These problems can be solved by introducing another two Higgs doublets and two singlets transforming as: $H^q_1: (1,2)(1/2, 1/3)$, $H^q_{2}: (1, 2)(1/2, -1/3)$,
$S_1: (1, 1)(0, 1)$ and $S_2: (1,1)(0, 2)$. The notation $(a,b)(c,d)$ refers to the representations under $SU(3)_C \times SU(2)_L \times U(1)_Y \times U(1)_{B_2-L_\mu}$.  $a$, $b$, $c$ and $d$ are the quantum numbers under $SU(3)_C$, $SU(2)_L$,  $U(1)_Y$ and  $U(1)_{B_2-L_\mu}$, respectively. In the following we will use the same notation. The following Yukawa couplings $\tilde L^q_{Y-mass}$ can be added to the Lagrangian
\begin{eqnarray}
\tilde L^q_{Y-mass} = -\left (\bar Q_L (Y^u_{H^q_1}  \tilde H^q_1 +Y^u_{H^q_2} \tilde H^q_2) U_R +
\bar Q_L (Y^d_{H^q_1}  H^q_1 +Y^d_{H^q_2} H^q_2) D_R
 + {1\over 2} \bar \nu_R^c (Y_{S_1} S_1 + Y_{S_2} S_2) \nu_R \right ) +  \mbox{H.c.}\;,
 \end{eqnarray}
 where
\begin{eqnarray}\label{yukawa}
&&Y_{H^q_1}^u = \left ( \begin{array}{ccc}
\;\;0\;\;&\;\;\;\;Y^{u12}_{H^q_1}\;\;\;\;&\;\;0\;\;\\
\;\;0\;\;&\;\;\;0\;\;&\;\;0\;\;\\
\;\;0\;\;&\;\;\;\;Y^{u32}_{H^q_1}\;\;\;\;&\;\;0\;\;
\end{array}
\right )\;,\;\;\;\;Y_{H^q_1}^d = \left ( \begin{array}{ccc}
\;0\;\;&\;\;0\;\;&\;\;\;0\;\;\\
\;Y^{d21}_{H^q_1}\;\;&\;\;0\;\;&\;\;\;Y^{d23}_{H^q_1}\;\;\\
\;0\;\;&\;\;0\;\;&\;\;\;0\;\;
\end{array}
\right )\;,\nonumber\\
&&Y_{H^q_2}^u = \left ( \begin{array}{ccc}
0\;\;&\;\;\;\;0\;\;&\;\;\;\;0\;\\
Y^{u21}_{H^q_2}\;\;&\;\;\;\;0\;\;&\;\;\;\;Y^{u23}_{H^q_2}\;\\
0\;\;&\;\;\;\;0\;\;&\;\;\;\;0\;
\end{array}
\right )\;,\;\;\;\;
Y_{H^q_2}^d = \left ( \begin{array}{ccc}
\;\;0\;\;&\;\;Y^{d12}_{H^q_2}\;\;&\;\;\;0\;\;\;\;\\
\;\;0\;\;&\;\;0\;\;&\;\;\;0\;\;\;\;\\
\;\;0\;\;&\;\;Y^{d32}_{H^q_2}\;\;&\;\;\;0\;\;\;\;
\end{array}
\right )\;,\\
&&Y_{S_1}\; = \left ( \begin{array}{ccc}
0\;\;&\;\;Y^{12}_{S_1}\;\;&\;\;0\\
Y^{12}_{S_1}\;\;&\;\;0\;\;&\;\;Y^{23}_{S_1}\;\\
0\;\;&\;\;Y^{23}_{S_1}\;\;&\;\;0
\end{array}
\right )\;,\;\;\;\;Y_{S_2}\; = \left ( \begin{array}{ccc}
\;\;0\;\;&\;\;\;0\;\;&\;\;\;0\;\;\;\;\\
\;\;0\;\;&\;\;\;Y^{22}_{S_2}\;\;&\;\;\;0\;\;\;\;\\
\;\;0\;\;&\;\;\;0\;\;&\;\;\;0\;\;\;\;
\end{array}
\right )\;.\nonumber
\end{eqnarray}

When Higgs bosons develop non-zero VEVs $v_i/\sqrt{2}$, the mass matrices $M_{u,d}$ for up and down quarks are given by
\begin{eqnarray}
M_u = {v_0\over \sqrt{2}} Y^u_{H} +  {v^q_1\over \sqrt{2}} Y^u_{H^q_1} + {v^q_2\over \sqrt{2}} Y^u_{H^q_2}\;,\;\;\;\;
M_d = {v_0\over \sqrt{2}} Y^d_{H} +  {v^q_1\over \sqrt{2}} Y^d_{H^q_1} + {v^q_2\over \sqrt{2}} Y^d_{H^q_2}\;,
\end{eqnarray}
the mass matrices $M_{l,\;\nu}$ for charged leptons and neutrinos in the basis $(\nu^c_L, \nu_R)^T$ are given by
\begin{eqnarray}
M_l = {v_0\over \sqrt{2}} Y^l_H\;,\;\;\;\;
M_\nu = \left ( \begin{array}{cc}
\;\;0\;\;&\;\;M_D\;\;\\
\;\;M_D^T\;\;&\;\;M_R\;\;
\end{array}
\right )\;,
\end{eqnarray}
where $M_D = (v_0/\sqrt{2}) Y_H^\nu$ and $M_R = \tilde M_R + (v_{S_1}/ \sqrt{2}) Y_{S_1} +  (v_{S_2}/ \sqrt{2}) Y_{S_2}$. We assume that the elements in $M_R$ are much larger than those in the other mass matrices so that the seesaw mechanism is  effective.

The above Yukawa couplings for charged leptons will not cause any mixing between different flavors of charged leptons so that $Z'$ only couples to $\mu$,  which leads to a vector-like coupling $Z'$ to muon as the desired structure to obtain a positive contribution to muon $g-2$ to solve the anomaly.  We need to introduce some Higgs doublets to have more involved mixing for neutrino phenomenology and also to reduce some potential difficulties for neutrino trident  data.
For this purpose, we can introduce a similar pair of Higgs doublets $H^l_{1,2}$, lepton counterparts of the quark scenario, with different charges $H_1^l: (1, 2) (1/2, 1)$ and $H^l_2: (1, 2) (1/2, -1)$   to allow the following Yukawa couplings
\begin{eqnarray}
\tilde L^l_{Y-mass} = -\left (\bar L_L (Y^\nu_{H^l_1}  \tilde H^l_1 +Y^\nu_{H^l_2} \tilde H^l_2) \nu_R +
\bar L_L (Y^l_{H^l_1}  H^l_1 +Y^l_{H^l_2} H^l_2 ) E_R \right ) + \mbox{H.c.}\;,
\end{eqnarray}
with
\begin{eqnarray}\label{yukawa}
&&Y_{H^l_1}^\nu = \left ( \begin{array}{ccc}
\;\;0\;\;&\;\;\;\;0\;\;\;\;&\;\;0\;\;\\
\;\;Y^{\nu21}_{H^l_1}\;\;&\;\;0\;\;&\;\;Y^{\nu23}_{H^l_1}\;\;\\
\;\;0\;\;&\;\;\;\;0\;\;\;\;&\;\;0\;\;
\end{array}
\right )\;,\;\;\;\;Y_{H^l_1}^l = \left ( \begin{array}{ccc}
\;\;\;0\;\;&\;\;Y^{l12}_{H^l_1}\;\;&\;\;\;0\;\;\;\\
\;\;\;0\;\;&\;\;0\;\;&\;\;\;0\;\;\;\\
\;\;\;0\;\;&\;\;Y^{l32}_{H^l_1}\;\;&\;\;\;0\;\;\;
\end{array}
\right )\;,\nonumber\\
&&Y_{H^l_2}^\nu = \left ( \begin{array}{ccc}
\;\;\;\;\;0\;\;&\;\;\;\;Y^{\nu12}_{H^l_2}\;\;&\;\;\;\;0\;\;\;\;\\
\;\;\;\;\;0\;\;&\;\;\;\;0\;\;&\;\;\;\;0\;\;\;\;\\
\;\;\;\;\;0\;\;&\;\;\;\;Y^{\nu32}_{H^l_2}\;\;&\;\;\;\;0\;\;\;\;
\end{array}
\right )\;,\;\;\;\;
Y_{H^l_2}^l = \left ( \begin{array}{ccc}
\;0\;\;&\;\;0\;\;&\;\;\;0\;\;\\
\;Y^{l21}_{H^l_2}\;\;&\;\;0\;\;&\;\;\;Y^{l23}_{H^l_2}\;\;\\
\;0\;\;&\;\;0\;\;&\;\;\;0\;\;
\end{array}
\right )\;,
\end{eqnarray}
and the mass matrices $M_l$ and $M_D$ are modified to
\begin{eqnarray}
M_l = {v_0\over \sqrt{2}} Y^l_{H} +  {v^l_1\over \sqrt{2}} Y^l_{H^l_1} + {v^l_2\over \sqrt{2}} Y^l_{H^l_2}
\;,\;\;\;\;
M_D = {v_0\over \sqrt{2}} Y^\nu_{H} +  {v^l_1\over \sqrt{2}} Y^\nu_{H^l_1} + {v^l_2\over \sqrt{2}} Y^\nu_{H^l_2}\;.
\end{eqnarray}
The above mass matrices will allow flavor changing $Z'$ interactions with charged leptons in the mass eigen-basis.

The above mass matrices $M_{u, d, l, D, R}$ are full $3\times 3$, and  $M_R$ is further full  symmetric matrix. We should diagonalize them by bi-unitary transformation
\begin{eqnarray}
M_f = V^{f\dagger}_L \hat M_f V_R^f\;,\;\; M_\nu = V^{\nu\; T} \hat M_\nu V^\nu\;.
\end{eqnarray}
Here $V^f_{L,R}$ are $3\times 3$ unitary matrices and $V^\nu$ is a $6\times 6$ unitary matrix whose left top corner $3\times 3$ matrix $V^\nu_{3\times 3}$ is an approximate unitary matrix assuming seesaw mechanism is  effective.
After fermion-mass diagonalization, KM and PMNS matrix can be accommodated as  $V_{KM} = V^u_L V^{d\dagger}_L$ and $V_{PMNS} \approx V^l_L V^{\nu\;T}_{3\times 3}$.

The SM-like Higgs boson $h$ will be dominated by the linear combination
$h = (v_0 h_H + v^q_1 h^q_1 + v^q_2 h^q_2 + v^l_1 h^l_1 + v^l_2 h^l_2)/ \sqrt{v_0^2+(v^q_1)^2+(v^q_2)^2 + (v^l_1)^2 + (v^l_2)^2}$.
Here $h_i$ are the real neutral component of each $H_i$. There are additional orthogonal combinations for the these real neutral fields. Correspondingly, there exist also other pseudoscalar bosons and  charged Higgs bosons. We will assume these new degrees of freedom are much heavier so that their effects are small. With this assumption, the new physics effects on SM particles will be dominated by  $Z'$ interaction terms.

Since $H^{q,l}_{1,2}$ have both $U(1)_{B_2-L_\mu}$ charges, there is in principle mixing between $Z$ and $Z'$. The mass-squared matrix in the basis $(Z,\;Z')^T$ is given by
\begin{eqnarray}
\left (\begin{array}{cc}
\;\;{g^2+ g^{\prime 2}\over 4} (v^2_0 + (v^q_1)^2 + (v^q_2)^2 + (v^l_1)^2 + (v^l_2)^2 )&\;\;{\tilde g\sqrt{g^2+g^{\prime 2}}\over 2} \left({1\over 3}((v^q_1)^2 - (v^q_2)^2)
+ (v^l_1)^2 - (v^l_2)^2\right)\;\;\\
\\
\;\;{\tilde g \sqrt{g^2 + g^{\prime 2}} \over 2}\left({1\over 3}((v^q_1)^2 - (v^q_2)^2)
+ (v^l_1)^2 - (v^l_2)^2\right)\;\;&\;\;\tilde g^2 ({1\over (3)^2}((v^q_1)^2+(v^q_2)^2) + (v^l_1)^2+(v^l_2)^2 + v^2_{S_1} + 4 v^2_{S_2})
\end{array}
\right )\;.
\end{eqnarray}
Note that $Z'$ mass can be much larger or smaller than $Z$ mass. In the large $Z'$ mass limit, the mixing between $Z$ and $Z'$ is of order $m_Z^2/m_{Z'}^2$.  If setting $v^q_1=v^q_2$ and $v^l_1=v^l_2$, the mixing is eliminated so that $m_Z^2 = (g^2+ g^{\prime 2}) (v^2_0 + (v^q_1)^2 + (v^q_2)^2 + (v^l_1)^2 + (v^l_2)^2 )/4$ and $m_Z'^2=\tilde g^2 (((v^q_1)^2+(v^q_2)^2)/(3)^2 + (v^l_1)^2+(v^l_2)^2 + v^2_{S_1} + 4 v^2_{S_2})$. To reduce the parameters in the numerical analysis, we will make the above choice.
In this case,
the relevant interactions of $Z'$ with quarks and charged leptons in the mass eigen-state basis are given by
\begin{eqnarray}
L_{int-f} &=& {1\over 3} \tilde g (\bar U_L V^u_L N_qV_L^{u\dagger} \gamma^\mu U_L + \bar U_R V^u_R N_qV_R^{u\dagger} \gamma^\mu U_R   + \bar D_L V^d_L N_q V_L^{d\dagger} \gamma^\mu D_L + \bar D_R V^d_R N_q V_R^{d\dagger} \gamma^\mu D_R) Z'_\mu\nonumber\\
&-& \tilde g (\bar l_L V^l_L N_l V_L^{l\dagger} \gamma^\mu l_L + \bar l_R V^l_R N_l V_R^{l\dagger} \gamma^\mu l_R ) Z'_\mu\;, \label{z-p-int}
\end{eqnarray}
where $N_q =N_2$ is the same as $N_l=N_\mu$ which is a diagonal matrix $diag \;(0,\;1,\;0)$.

From the structure of mass matrices, we find that the unitary matrices $V^{u,d}_{L,R} = (V^{u,d}_{L,R\; ij})$, in general, have all non-zero entries. $V_L$ and $V_R$ are generally different.
Phenomenologically a vector-like $Z'$ coupling to leptons is favored because it provides a positive contribution to muon $g-2$.  We also have in mind to use $Z'$ to produce the required interaction for addressing the $b\to s \mu \bar \mu$ induced $B$ anomaly which favors $\bar s \gamma^\mu L b Z^\prime_\mu$ type of coupling.
We will use the following sub-set of the  $Z^\prime$-fermion interaction contained in the above interaction Lagrangian  for our detailed studies,
\begin{eqnarray}
L_{int-sb, \mu,\tau} = &&- \tilde g \left (( V^l_{L\;22} V^{l*}_{L\;22}) \bar \mu \gamma^\mu \mu
+ ( V^l_{L\;32} V^{l*}_{L\;32}) \bar \tau \gamma^\mu \tau + ( V^l_{L\;22} V^{l*}_{L\;32}) \bar \mu \gamma^\mu \tau
+ ( V^l_{L\;32} V^{l*}_{L\;22}) \bar \tau \gamma^\mu \mu \right )Z'_\mu \nonumber\\
&&+ {1\over 3} \tilde g \left (V^d_{L\; 22} V^{d*}_{L\;32} \bar s_L \gamma^\mu b_L
\right ) Z'_\mu\;. \label{zprime-current}
\end{eqnarray}
\\

\section{ $U(1)_{B_2-L_\mu}$ and the  muon $(g-2)_\mu$ anomaly}

\noindent
{\bf Contribution to muon $(g-2)_\mu$}

\begin{figure}[!t]
	\centering
	{\includegraphics[width=.486\textwidth]{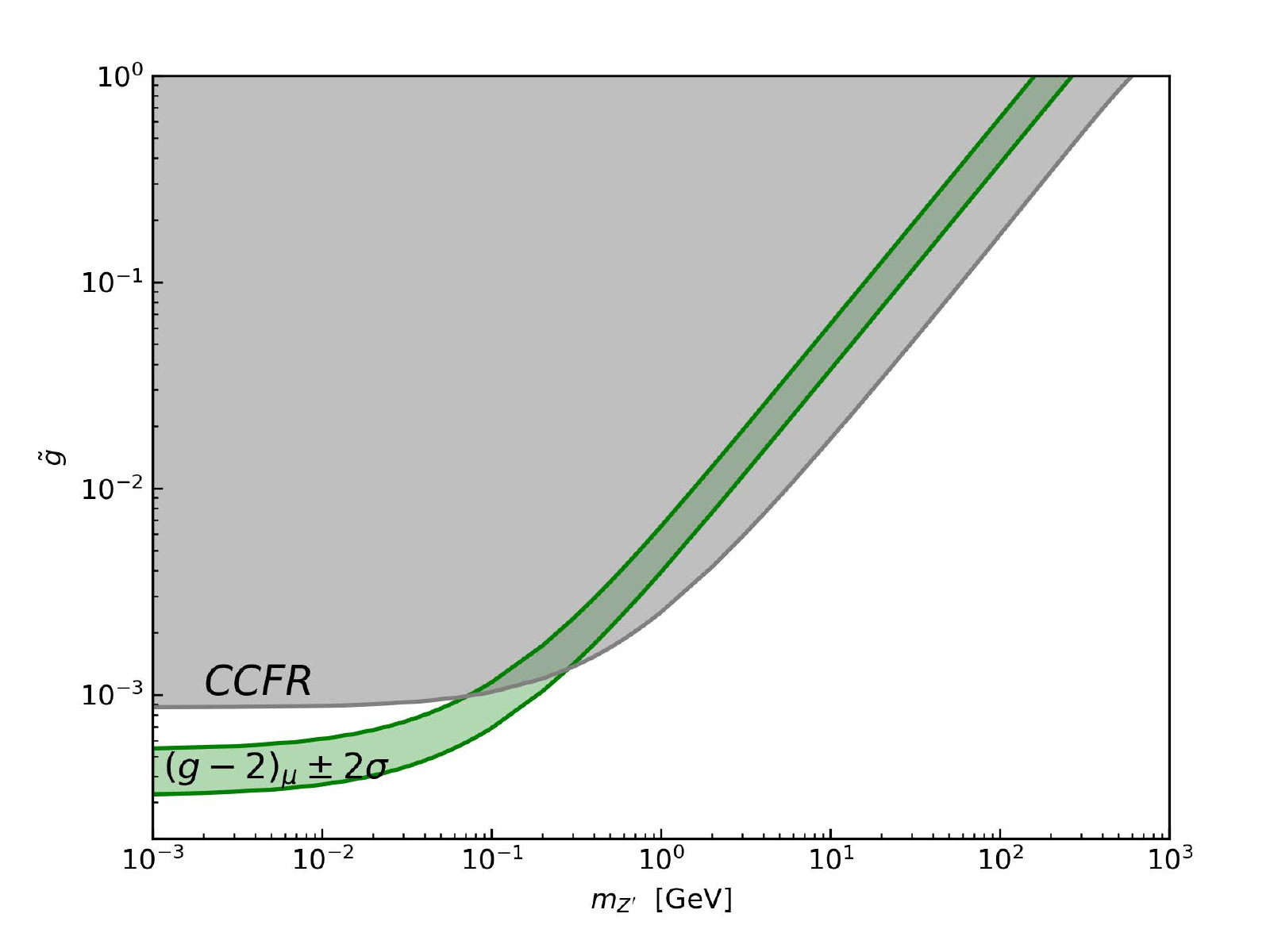}}
	\caption{The $(g-2)_\mu$ allowed region in $\tilde g - m_{Z'}$ plane with $V^l_{L\;22} = 1$. The excluded region by neutrino trident process is in gray.
}
	\label{g_2Trident}
\end{figure}

Using the $Z'$ interaction with charged leptons, we obtain the contribution to muon $g-2$ at one loop level to be~\cite{Leveille:1977rc}
\begin{eqnarray}
\Delta a_\mu^{new}  =\frac{\tilde{g}^{2} m_{\mu}^{2}}{4 \pi^{2}}\left(V^l_{L\; 22} V_{L \;22}^{l*}\right)
\int_{0}^{1} d x  \left[  \frac{(V_{L\;32}^l V_{L\;3 2}^{l*}) \left(x-x^{2}\right)\left(x+\frac{2 m_{\tau}}{m_{\mu}}-2\right)}{m_{\mu}^{2} x^{2}+m_{Z^{\prime}}^{2}(1 - x)+\left(m_{\tau}^{2}-m_{\mu}^{2}\right)x}
 +  {\left(V^l_{L\;2 2} V_{L\; 2 2}^{l*}\right) x^2(1-x)  \over  m^2_\mu x^2+m^2_{Z'}(1-x)} \right]\;. \label{a-anomaly}
\end{eqnarray}
In the limit $m_{Z'} >> m_{\tau,\mu}$,
\begin{equation}
	\Delta a^{new}_\mu = \frac{\tilde{g}^{2}}{4 \pi^{2}} \frac{m_{\mu}^{2}}{m_{Z^{\prime}}^{2}}\left(V^l_{L\; 22} V_{L\; 22}^{l*}\right)\left[\left(V^l_{L\; 32} V_{L\;32}^{l*}\right)
\left(\frac{m_{\tau}}{m_{\mu}}-\frac{2}{3}\right)+\frac{1}{3} \left(V^l_{L\; 22} V_{L\; 22}^{l*}\right)\right]\;,
\end{equation}
whereas for the opposite limit $m_{Z'}\leq m_\mu$, one should use the full integral form in Eq.~(\ref{a-anomaly}).
Assuming the difference $\Delta a_\mu = (251\pm 59)\times 10^{-11}$ is due to $\Delta a^{new}_\mu$ and using the relation $|V^l_{L\; 32}|^2=1-|V^l_{L\; 22}|^2$, we obtain the allowed parameter space by muon $(g-2)_\mu$ in the $\tilde g-m_{Z'}$ plane with the green narrow band   in Fig.~\ref{g_2Trident}.

The above show that as far as only $(g-2)_\mu$ problem is concerned, it is easy to deal with.
There are, however,  several processes which give strong constraints on the mass $m_{Z'}$ and coupling constant $\tilde g$, such as neutrino trident production~\cite{Altmannshofer:2014pba,Bonilla:2017lsq,Allanach:2020kss,Altmannshofer:2014cfa}, LHC $Z'$ searches~\cite{Bonilla:2017lsq,Allanach:2020kss} and Z coupling to leptons~\cite{Bonilla:2017lsq,Altmannshofer:2014cfa,Altmannshofer:2016brv}.
We will show below that experimental bounds from neutrino trident data will restrict $m_{Z'}$ to be less than a few hundred MeV if there is no $\mu-\tau$ mixing, that is, $|V^l_{L\;22}| =1$. Data from $\tau \to \mu \bar \mu \mu$, $\tau \to \mu \gamma$, however, only allow $|V^l_{L\;22}| =1$ to solve the $(g-2)_\mu$ problem. Therefore in this model the $Z^\prime$ mass is restricted to be less than a few hundred MeV.
\\

\noindent
{\bf Constraints from neutrino trident  and other processes}

The new $Z^{\prime}$ coupling to $\mu$ contributes to produce a $\mu^+\mu^-$
pair in the neutrino trident process, $\nu_{\mu} N \rightarrow \nu_{\mu} N \mu^{+} \mu^{-}$. This provides a sensitive probe for $\tilde g$. In the heavy $Z'$ limit, our model gives the correction to the cross section as
\begin{equation}
	\frac{\sigma^{NP}}{\sigma^{SM}}=1+\frac{\left(1+4 s_{W}^{2}\right) \frac{4 v^{2}{\tilde{g}}^{2}}{m_{Z^{\prime}}^{2}}(V^l_{L\; 22}V^{l*}_{L\; 22})+\frac{4 v^{4} \tilde{g}^{4}}{m_{Z^{\prime}}^{4}}(V^l_{L\; 22}V^{l*}_{L\; 22})^{2} }{1+\left(1+4 s_{W}^{2}\right)^{2}}\;,
	\label{trident}
\end{equation}
where  $v=246$ GeV and $s_{W}=\sin \theta_{W}$ with  weak Weinberg mixing angle $\theta_W$.

Using the experimental measurement  within $2\sigma$ error for trident cross section by the CCFR collaboration~\cite{Mishra:1991bv}, $\sigma^{\mathrm{CCFR}}/\sigma^{\mathrm{SM}}=0.82 \pm 0.28$, we obtain the excluded parameter space in the $\tilde{g}-m_{Z'}$ plane in gray as shown in Fig.~\ref{g_2Trident}.
Neutrino trident production is also observed by other experiments, such as CHARM-II~\cite{Geiregat:1990gz} and NuTeV~\cite{Adams:1999mn}. If combining these three collaborations, we get the global average value as $\sigma_{\exp } / \sigma_{\mathrm{SM}}=0.95 \pm 0.25$. It is slightly weaker than the bound from CCFR. So we use the more stringent CCFR bound to analyze the constraints in our model.  For a large $m_{Z^\prime}$, if there is no mixing between $\mu$ and $\tau$, $|V_{L\;22}^l |^2= 1$, $\sigma^{NP}/\sigma^{SM}$ is predicted to be larger than 5. Therefore in order to have a large $Z^\prime$ mass, one should choose a $|V_{L\;22}^l |^2$ which is much smaller than 1. This is why we introduced the mixing and we will try to see if this indeed can achieve providing a solution for a large $m_{Z^\prime}$. With $|V_{L\;22}^l |^2=1$, the model can still be made to accommodate the  neutrino trident data  to solve $(g-2)_\mu$ anomaly, but in this case $Z^\prime$ mass is restricted to be  less than a few hundred MeV as can be seen from Fig.~\ref{g_2Trident}.

We comment that for small $m_{Z'}$  with a few hundred MeV, the large $Z'$ mass approximation in Eq.~(\ref{trident}) is no longer accurate enough.  We have taken the  $q^2$ dependence  into account to obtain the constraints as shown in
Fig.~\ref{g_2Trident}  based on Ref.~\cite{Altmannshofer:2014pba}.  We find that there is allowed parameter space to accommodate the $(g-2)_\mu$ anomaly if the $Z'$ mass is below  300 MeV.
\\


\begin{figure}[!t]
	\centering
	\subfigure[\label{contrains_BIG_MZ}]
	{\includegraphics[width=.486\textwidth]{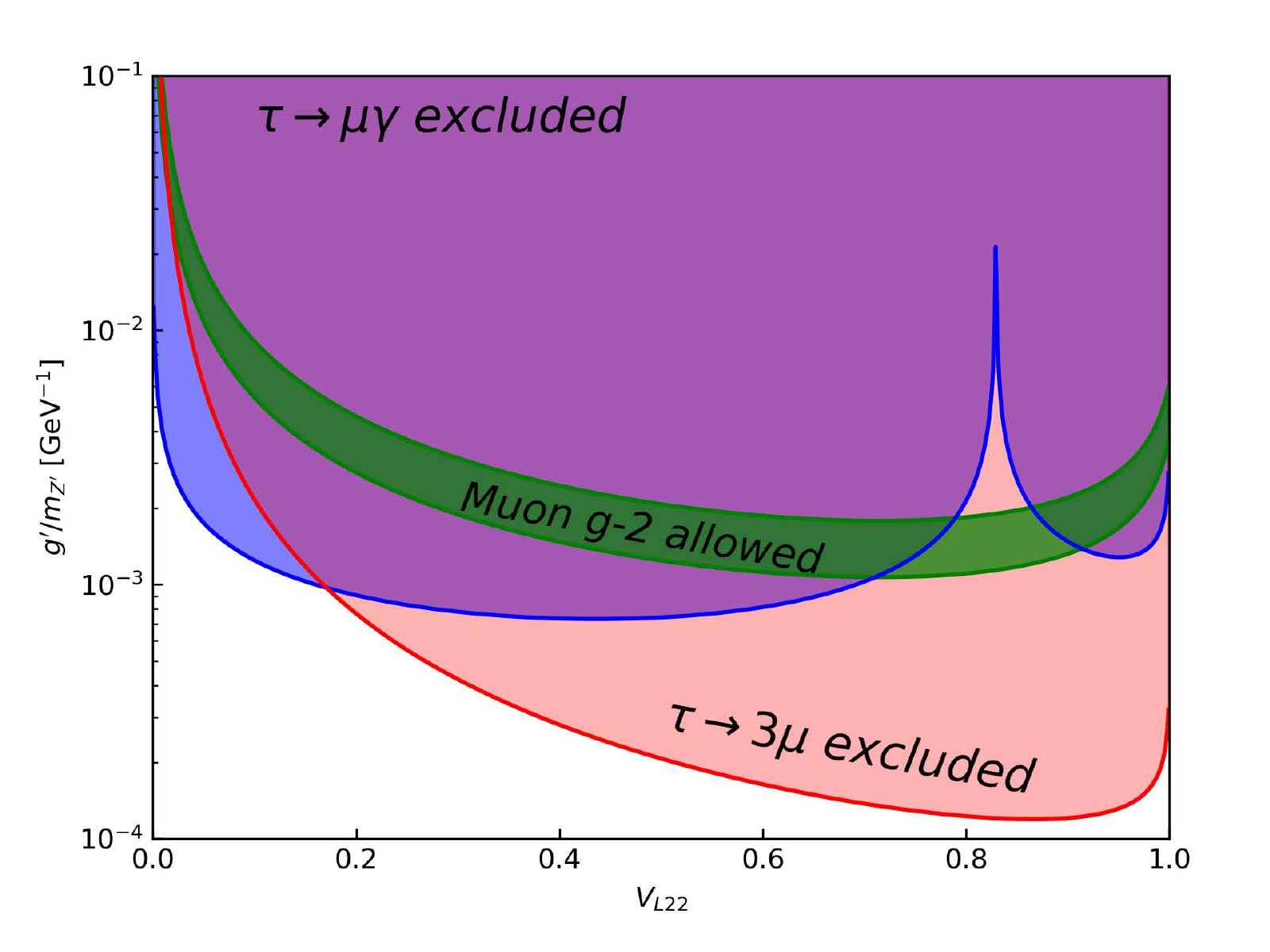}}
	\subfigure[\label{contrains_MZ_01}]
	{\includegraphics[width=.486\textwidth]{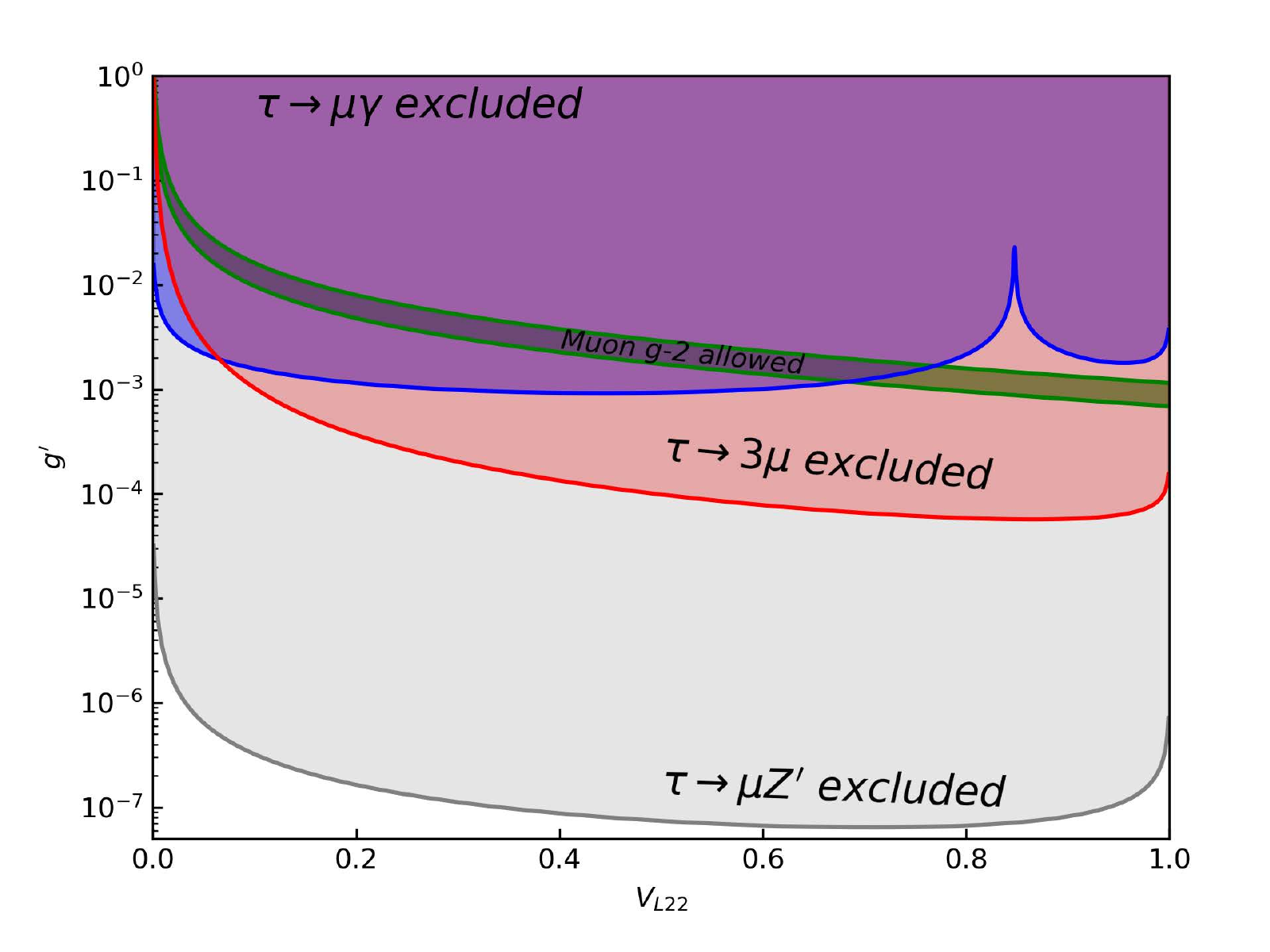}}
	\caption{  The excluded region in the large and small $m_{Z'}$ cases. The excluded region by $\tau \to 3\mu $ is in red, $\tau \to \mu \gamma$  in blue, $\tau \to \mu Z^{\prime}$ in gray, respectively. 		 The narrow band between green line is the allowed parameter space for explaining muon g-2 anomaly.
 (a) The left panel  is for large $m_{Z'}$. (b) The right panel is for small $m_{Z'}$ with $m_{Z'} = 0.1$ GeV.
	}
	\label{contrains_MZ1}
\end{figure}


\noindent
{\bf Constraints from $\tau \to \mu \bar \mu \mu, \;\mu \gamma$ and $\tau \to \mu Z'$}

We now study if  a $V^l_{L\; 22}$ deviating from 1, is allowed so that one has a chance to have a large $m_{Z^\prime}$ to solve $(g-2)_\mu$ anomaly. If $V_{L\;32}^l*V^l_{L\;22}$ is not zero, $\tau \to \mu \bar \mu \mu$ and $\tau \to \mu \gamma$ will occur at the tree level and the one loop level, respectively. And further if $m_{Z'}<m_\tau - m_\mu$, $\tau \to \mu Z'$ is kinematically allowed to happen. The decay amplitudes are given by
\begin{eqnarray}
	&&M(\tau \to \mu \bar\mu \mu) = \tilde g^2 |V^l_{L\;22}|^2V_{L\;22}^l V^{l*}_{L\;32}\bar \mu(p_3) \gamma^\alpha  \mu(p_4) {1 \over (p_3+p_4)^2 - m^2_{Z'}} \bar \mu(p_2) \gamma_\alpha  \tau(p_\tau)-(p_2 \leftrightarrow p_3)\;,\nonumber\\
	&&M(\tau \to \mu Z') = \tilde g V^l_{L\;22}V^{l*}_{L\;32} \epsilon^{*\alpha}(Z') \bar \mu \gamma_\alpha \tau\;,
\end{eqnarray}
and the interaction Lagrangian inducing $\tau \to \mu \gamma$ at one loop is
\begin{eqnarray}
\mathcal{L}_{NP} =e m_\tau C_L^{\mu \tau} \bar l_\mu \sigma_{\alpha\beta} P_L l_\tau F^{\alpha\beta}+ e m_\tau C_R^{\mu \tau} \bar l_\mu \sigma_{\alpha\beta} P_R l_\tau F^{\alpha\beta}\;.
\end{eqnarray}
Here
\begin{equation}
\label{intergaltaugamma}
 C_L^{\mu \tau}= C_R^{\mu \tau}= \sum_k {C^{\mu k}C^{k \tau}\over 16\pi^2 m_\tau}
\int_{0}^{1} d x d y \frac{x\left[(y-1)m_\mu-(x+y)m_\tau\right]-2m_k(2y-1)}{x m_{Z^{\prime}}^{2}+(1-x)m_k^2-x(1-x-y)m_\tau^2-x y m_\mu^2}\;,
\end{equation}
where $k$ can be $\mu$ or $\tau$  circulated in the loop. $C^{\mu k}$ and $C^{k \tau}$ are the coupling vertexes in our model for charged lepton to $Z^{\prime}$ with the forms $C^{\mu \mu} = -\tilde{g} V^l_{L\;22} V^{l*}_{L\;22}$, $C^{\mu \tau} = -\tilde{g} V^l_{L\;22} V^{l*}_{L\;32}$ and $C^{\tau \tau} = -\tilde{g} V^l_{L\;32} V^{l*}_{L\;32}$ .

Based on the above amplitudes, we can obtain the relevant branching ratios. Using experimental data $Br^{exp}(\tau \to 3\mu) < 2.1 \times 10^{-8}$, $Br^{exp}(\tau \to \mu \gamma) < 4.4 \times 10^{-8}$ in 90$\%$ C.L. and $Br^{exp}(\tau \to \mu Z')< 5\times 10^{-3}$~\cite{Zyla:2020zbs}, we obtain the excluded parameter space for large $m_{Z'}$ in Fig.~\ref{contrains_BIG_MZ} and small $m_{Z'}$ in Fig.~\ref{contrains_MZ_01}, respectively.
We find that the strongest constraints are from $\tau\to 3\mu$ and $\tau\to \mu Z'$, courtesy of large and
small $m_{Z'}$ cases. And for whichever case (large or small $m_{Z'}$), the muon $(g-2)_\mu$ allowed region in green has been ruled out.
Therefore, we find that $|V^l_{L\;22}| \sim 1 \;(|V^l_{L\;23}|\sim 0)$ is needed to forbid to exist flavor changing $\tau$ decay processes while solving $(g-2)_\mu$ problem.
Similarly, non-zero factor $V^l_{L\;22}*V^{l*}_{L\;32}$ will also induce Michel decay process $\tau \to \mu \bar \nu \nu$ by exchanging $Z'$.  
The Michel parameters will further be influenced. However, this constraint is weaker compared with the ones from $\tau\to 3\mu$ and $\tau \to \mu \gamma$.
\\

We now comment on several other constraints.
Di-muon pair production process  $pp\to Z'\to \mu^+\mu^-$ gives the strongest constraint on the model parameters at large $m_{Z^{\prime}}$ region. The ATLAS and CMS experiments have performed various searches in $pp$ collisions at the LHC for
resonant $Z'$ vector bosons decaying into different final states. Due to the null signal of di-muon resonance up to date, the lower
limits are placed upon the production cross sections times branching ratio as a function of the invariant mass
of the final state. The null result signal of di-muon can also constrain the  parameter space in our model by using ATLAS~\cite{Aaboud:2017buh} with $150\;\mbox{GeV} \leq m_{Z'} \leq 5\;\mbox{TeV}$ and CMS~\cite{Sirunyan:2018exx} with
$200\;\mbox{GeV} \leq m_{Z'} \leq 5.5\;\mbox{TeV}$. For MeV $m_{Z^{\prime}}$ region, the constraints from di-muon resonance will not apply again.

Lepton flavor universality (LFU) of  Z couplings can provide another set of constraints for the parameters.  The presence of $Z'-\mu\mu$ and $Z'-\nu \nu$ couplings will break LFU in Z boson decay. This is manifest in Z couplings to muons and neutrinos through loop effects. The corrections to the vector and axial vector couplings of $Z\mu\mu$ relative to the Standard-Model-like $Z-ee$ can be expressed as
\begin{eqnarray}
{g_{V\mu}\over g_{Ve}} \simeq  {g_{A\mu}\over g_{Ae}}\simeq \left| 1+{\tilde g^2  \over (4\pi)^2} (V^{l}_{L\;22} V_{L\;22}^{l*}) k_F(m_Z^2/m_{Z'}^2)\right|\;.
\label{Zmumu}
\end{eqnarray}
And similarly for $Z \nu \nu$, out of the three SM neutrinos only the muon-neutrino in the weak basis is  affected by $Z'$ loops. Therefore, the
correction to Z coupling to neutrino is effectively given by
\begin{eqnarray}
{g_{V\nu}\over g_{Ve}} \simeq  {g_{A\nu}\over g_{Ae}}\simeq \left|1+{\tilde g^2  \over (4\pi)^2}  {1\over 3} k_F(m_Z^2/m_{Z'}^2)\right|\;.
\label{Znunu}
\end{eqnarray}
Here $k_F$ is the loop factor that can be found in Ref.~\cite{Haisch:2011up}. The vector and axial vector
couplings of Z boson  can be found from electroweak measurements in Ref.~\cite{Zyla:2020zbs}.  The relevant ones are $g_{Ve}=-0.03817\pm 0.00047$, $g_{Ae}=-0.50111\pm 0.00035$, $g_{V\mu}=-0.0367\pm 0.0023$, $g_{A\mu}=-0.50120\pm 0.00054$, $g_{V\nu}=g_{A\nu}=0.5008\pm 0.0008$. We find that the most stringent constraint is  from $g_{A\mu}/g_{Ae}=1.00018\pm 0.00128$ which is consistent with Ref.~\cite{Bonilla:2017lsq}. For $m_{Z'}$ less than a few hundred MeV, these constraints are safely satisfied.

We conclude that the $U(1)_{B_2-L_\mu}$ model discussed here is able to solve the $(g-2)_\mu$ problem if the $Z'$ mass is less than a few hundred MeV.
\\

\section{$U(1)_{B_2-L_\mu}$ and $b\to s \mu \bar \mu$ induced $B$ anomalies}

We now study  $b\to s \mu \bar \mu$ induced anomalies in $U(1)_{B_2 - L_\mu}$ model.
Using Eq.~(\ref{zprime-current}), by exchanging $Z'$ at tree level the required effective Hamiltonian will be generated with
\begin{eqnarray}
	H_{eff} &&= {\tilde g^2(g^{\alpha\beta} - q^\alpha q^\beta/m^2_{Z'}) \over 3(q^2 - m^2_{Z'})} \left(V^d_{L\;22}V^{d*}_{L\;32}\right)\left(V^l_{L\;22}V^{l*}_{L\;22}\right) \left (\bar s_L \gamma_\alpha b_L \bar \mu \gamma_\beta \mu\right )\nonumber\\
	&& = {\tilde g^2 \over 3(q^2 - m^2_{Z'})} \left(V^d_{L\;22}V^{d*}_{L\;32}\right)\left(V^l_{L\;22}V^{l*}_{L\;22}\right) \left (\bar s_L \gamma_\alpha b_L \bar \mu \gamma^\alpha \mu\right )\;, \label{c9exp}
\end{eqnarray}
where $q$ is the momentum transfer from quarks to muons. For the last line in the above equation, we  use $\bar \mu \gamma^\alpha q_\alpha \mu =0$.

\begin{figure}[!t]
	\centering
	\subfigure[\label{C9DDmixingcons}]
	{\includegraphics[width=.486\textwidth]{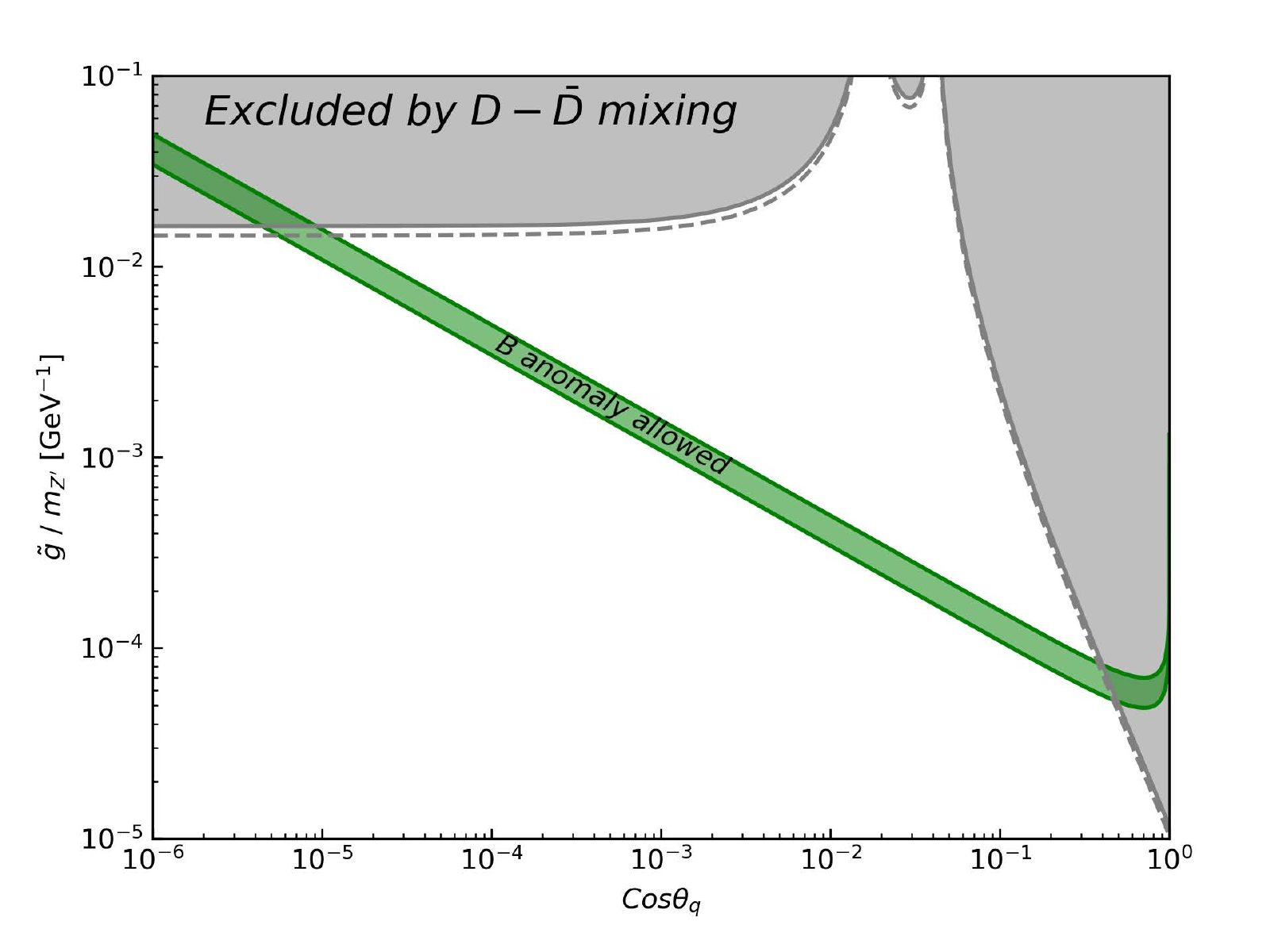}}
	\subfigure[\label{C9VL22100}]
	{\includegraphics[width=.486\textwidth]{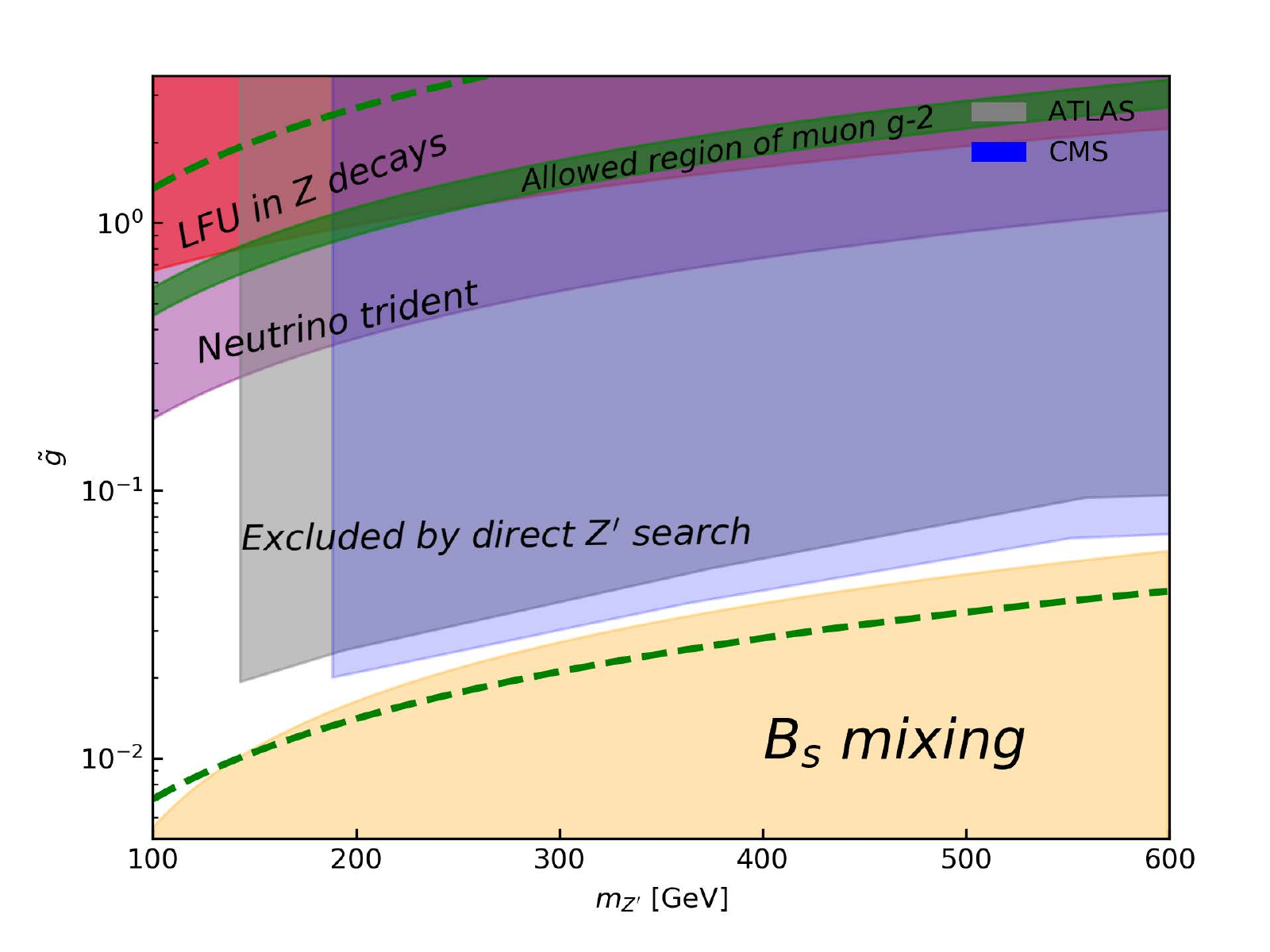}}
	\caption{ (a) The allowed parameter space in the $\tilde g/m_{Z'}-\cos \theta_q$ plane for large $m_{Z^{\prime}}$. The narrow green band is the allowed region to explain $b\to s \mu\mu$ induced anomalies. The gray region is excluded by $D-\bar{D}$ mixing by choosing $c(m_{Z^\prime}) = 0.8$ and the dashed curve is the exclusion line by choosing $c(m_{Z^\prime})= 1$.
	(b) The excluded region in the $\tilde g-m_{Z'}$ plane for large $m_{Z^{\prime}}$. The excluded region by neutrino trident production is in purple,  LHC $Z'$ search in gray, LFU in $Z$ decay in red, $B_s$ mixing in yellow, respectively. The narrow green band is the allowed parameter space for explaining muon g-2 anomaly. The region between the two green dashed line is the explanation of the $B$ anomalies combined with the constraint of $D$ mixing in fact, and the yellow area is excluded by  $B_s$ mixing combined with the $B$ anomalies explanation. 
	}
	\label{constrainsV2LL}
\end{figure}

Writing the above into the standard form $H_{eff} = - (G_F\alpha_{em}/ \sqrt{2} \pi) V_{tb} V^*_{ts} \sum_i C_i O_i$, we have the following operator and corresponding coefficient in the $m^2_{Z'} >> q^2$ limit
\begin{eqnarray}
O_9 = \bar s_L\gamma^\mu b_L \bar \mu \gamma_\mu \mu\;,\;\;\;\;\;\;\;\;\;\;\;\;\;\;\;
C^{new}_9 = {\sqrt{2}\pi\tilde g^2 \over 3 m^2_{Z'}G_F \alpha_{em}} \left(V^l_{L\; 22} V_{L\; 22}^{l*}\right) {V_{L\; 22}^d V^{d*}_{L\;32}\over V_{tb}V^*_{ts}}\;.
\end{eqnarray}
It has been shown that the contribution, dominated by $C_9 = -0.80\pm 0.14$  ~\cite{review-fit} with 5.7$\sigma$ 
pull from SM best fit, is the best scenario to explain $b\to s \bar \mu \mu$ anomalies from global rare $B$ decay data fit.
Our model can naturally accommodate this scenario by setting $V^{d}_{R}$ to be zero.
We would obtain
\begin{eqnarray}
\frac{\tilde{g}^2}{m^2_{Z^{\prime}}} {V^d_{L\; 22}V^{d*}_{L\;32}\over V_{tb} V_{ts}^*} =
 {3G_F \alpha_{em} \over \sqrt{2}\pi} {C_9 \over  |V^l_{L\; 22}|^2} \;.\label{C9constrains}
\end{eqnarray}
To produce the required $C_9$, we have $(\tilde{g}^2/m^2_{Z^{\prime}}) |V^l_{L\;22}|^2 (V^d_{L\; 22}V^{d*}_{L\;32}/ V_{tb} V_{ts}^*) = (- 0.46\pm 0.08) \times 10^{-7}\;\mbox{GeV}^{-2}$.
The allowed parameter region  within $2\sigma$ error to explain $b\to s \bar \mu \mu$ anomalies is shown in $\tilde g/m_{Z'}-\cos \theta_q$ plane with green in Fig.~\ref{C9DDmixingcons}. Here $\theta_q$  means the down type of quarks mixing angle with $\cos \theta_q=V^d_{L\;22}$ and $-\sin \theta_q=V^d_{L\;32}$.

To establish a solution for $b\to s \mu \bar \mu$ anomalies, one must make sure that known constraints from other processes are satisfied.
Besides these processes  discussed in the Sec. III,  $B\to K\nu \bar \nu$, $D\to \mu^+\mu^-$, $D$ and $B$ meson mixings will also  constrain the model parameters. In the following we provide some more information.

Firstly we comment on  $B\to K\nu \bar \nu$ and $D\to \mu^+\mu^-$. For $B\to K\nu \bar \nu$, this decay process is similar to $B\to K\mu \bar \mu$ just replacing $\mu$ by $\nu$. And for $D\to \mu^+\mu^-$, $Z'$ interaction induces the second generation of quarks transition to the first generation of quarks, such as $Z' -c \to u$ interaction,  which may constrain our model parameters by involving  the mixing matrix $V^{u}_{L}$, which are fully determined by $V_{CKM}$ and $V_L^{d}$~\cite{Alonso:2017uky},
\begin{eqnarray}
	V_L^u= V_{CKM} V_L^d
	&=&
	\left ( \begin{array}{ccc}
		V_{ud}\;\;&\;\;V_{us}\;\;&\;\;V_{ub}\\
		V_{cd}\;\;&\;\;V_{cs}\;\;&\;\;V_{cb}\\
		V_{td}\;\;&\;\;V_{ts}\;\;&\;\;V_{tb}
	\end{array}
	\right )
	\left ( \begin{array}{ccc}
		1\;\;&\;\;0\;\;&\;\;0\\
		0\;\;&\;\;\cos \theta_q\;\;&\;\;\sin \theta_q\\
		0\;\;&\;\;-\sin \theta_q\;\;&\;\;\cos \theta_q
	\end{array}
	\right )\nonumber\\
	&=&\left ( \begin{array}{ccc}
		V_{ud}\;\;&\;\;V_{us}\cos \theta_q-V_{ub}\sin \theta_q\;\;&\;\;V_{us}\sin \theta_q+V_{ub}\cos \theta_q\\
		V_{cd}\;\;&\;\;V_{cs}\cos \theta_q-V_{cb}\sin \theta_q\;\;&\;\;V_{cs}\sin \theta_q+V_{cb}\cos \theta_q\\
		V_{td}\;\;&\;\;V_{ts}\cos \theta_q-V_{tb}\sin \theta_q\;\;&\;\;V_{ts}\sin \theta_q+V_{tb}\cos \theta_q
	\end{array}
	\right ).
	\label{Dmixing}
\end{eqnarray}
Exchanging  $Z'$ at tree level, we have
	\begin{eqnarray}
		&&H_{B\to K \nu \bar\nu} =\frac{\tilde g^2}{3(q^2-m_{Z'}^2)} (V^d_{L\;22}V^{d*}_{L\;32})
		(\bar s_L \gamma_\mu b_L \bar \nu \gamma^\mu L \nu )\;,  \nonumber\\
		&&H_{c\to u \mu^+\mu^-} =\frac{\tilde g^2}{3(q^2-m_{Z'}^2)} (V^u_{L\;22}V^{u*}_{L\;22})|V^l_{L\;22}|^2
		(\bar u_L \gamma_\mu c_L \bar \mu \gamma^\mu  \mu)\;.
	\end{eqnarray}
Following  Ref.~\cite{He:2021yoz}, we find that the branching fraction for $B\to K\nu \bar \nu$ is $Br(B\to K \nu \bar\nu)\times 10^6 \approx4.39 -0.457 \mbox{Re} \sum_i \left(C_9/2+C^{SM}_L\right)^{ii}$. 
Here  $C_L^{SM}=-X(x_t)/s_W^2$ with $X(x_t)=1.469\pm0.017$~\cite{Buchalla:1998ba,Brod:2010hi}.
And the branching ratio for $D\to \mu^+\mu^-$ is zero  due to the vector current type of coupling form $\bar \mu \gamma^\mu \mu$~\cite{Golowich:2009ii}.
	
	In our model, we find for  $B\to K\nu \bar \nu$, by combining the experimental data $Br(B^+\to K^+\nu \bar \nu)=(1.1\pm0.4)\times 10^{-5}$~\cite{Dattola:2021cmw} and  SM prediction $Br(B^+\to K^+\nu \bar \nu)_{SM}=(4.4\pm0.7)\times 10^{-6}$, one obtains $R_K^\nu=Br(B^+\to K^+\nu \bar \nu)/Br(B^+\to K^+\nu \bar \nu)_{SM}=2.5\pm1.0$. 
	By using the central value of input parameters in our model, one obtains $R_K^\nu=1.01$ satisfying the constraint from experimental data and SM prediction.

Now we comment on $D-\bar D$ and $B_s-\bar B_s$ mixing. For $D-\bar D$ mixing, 
using  Eq.~(\ref{Dmixing}), we find the  contribution to $D-\bar D$ mixing matrix elements by exchanging $Z^{\prime}$ as
\begin{eqnarray}
\label{DDbarmixing}
-iM &&= -\tilde{g}^{2}(V^u_{L\;12} V^{u*}_{L\;22})^2
\left(\bar{u}_{L} \gamma_{\mu} c_{L} \frac{g^{\mu \nu}-q^{\mu}q^{\nu}/m^2_{Z^\prime}}{m^2_{c}-m^2_{Z^{\prime}}} \bar{u} \gamma_{\nu} c_L\right)\nonumber\\
&&= -\tilde{g}^{2}(V^u_{L\;12} V^{u*}_{L\;22})^2 \frac{1}{m^2_{c}-m^2_{Z^{\prime}}}
\left[(\bar{u}_{L} \gamma^{\mu} c_{L})( \bar{u}_L \gamma_{\mu} c_L)
-\frac{m_c^2}{m_{Z'}^2}(\bar{u}P_{R} c)(  \bar{u} P_R c)\right].
\end{eqnarray}
For $m_{Z'} >> m_{c}$, the second term in the bracket can be neglected and
this leads to the following mass difference
\begin{eqnarray}
\Delta m_D={2\over 3}f_D^2 B_D m_D c(m_{Z^\prime}){\tilde g^2 \over 18m^2_{Z^\prime}}\left[(V_{us} \cos \theta_q-V_{ub}\sin \theta_q)(V^*_{cs} \cos \theta_q-V^*_{cb}\sin \theta_q)\right]^2\;.\label{DD}
\end{eqnarray}
Here  $f_D= 0.2074$GeV~\cite{Carrasco:2014poa}, $B_D=0.757$~\cite{Carrasco:2015pra} and 
the Wilson coefficient $c(m_{Z^\prime})\approx 0.8$ includes the NLO running from electroweak scale down to the meson mass decay~\cite{Alonso:2017uky,Ciuchini:1997bw,Buras:2000if}. It becomes larger  when lowering $m_{Z'}$ and in the sub GeV scale~\cite{Buras:1984pq} is close to $c(m_{Z^\prime})\approx 1$. This leads to a stronger constraint for smaller $m_{Z'}$ as shown in Fig.~\ref{C9DDmixingcons} by the dashed curve.
Using  the experimental data
$\Delta m_D=(0.95^{+0.41}_{-0.44})\times 10^{10} s^{-1}$~\cite{Zyla:2020zbs}, we obtain the excluded region with $2\sigma$ error in gray shown in Fig.~\ref{C9DDmixingcons}. We find that there exist some suitable spaces  with $\cos \theta_q =(10^{-5},\;0.4)$ to explain the $B$ anomalies while satisfying the $D-\bar D$ mixing bound.
  The long-distance contribution within  SM can also contribute to the $D - \bar{D}$ mixing. 
The theoretical prediction on $D - \bar{D}$ mixing is model-dependent, which may lead to  large uncertainties~\cite{HFLAV:2019otj,Cheng:2010rv,Umeeda:2021llf,Jiang:2017zwr}. 
In our work, we only focus on  a rough estimate on $D-\bar{D}$ mixing from $Z'$ contribution for the short-distance part.  Therefore, here we have not considered the long-distance contributions and concentrated on the new short-distance ones.

For these processes with $V^l_{L\;22}=1$ discussed in Sec. III, we obtain the corresponding constraints within $2\sigma$ error as shown in Fig.~\ref{C9VL22100}. We find that although the most region has been ruled out by these processes, there is  suitable region to explain the $B$ anomalies.

A non-zero value $V^d_{L\; 22}V^{d*}_{L\;32}$ will induce $B_s-\bar{B_s}$ mixing by exchanging $Z'$ at the tree level with the amplitude by
\begin{eqnarray}
	M_{12} &&= M^{SM}_{12} +\left(\frac{1}{3} \tilde{g} V_{L\; 22}^{d} V_{L \;32}^{d *}\right)^{2}   {(g^{\alpha\beta} - q^\alpha q^\beta/m^2_{Z'})\over m^2_{Z'} - q^2} (\bar s \gamma_\alpha L b)( \bar s \gamma_\beta L b)  \nonumber\\
	&&=M^{SM}_{12} + \left(\frac{1}{3} \tilde{g} V_{L\; 22}^{d} V_{L\; 32}^{d *}\right)^{2} \frac{1}{m_{Z^{\prime}}^{2}-m^2_{B_s}}\left[\left(\bar{s} \gamma^{\alpha} L b\right)\left(\bar{s} \gamma_{\alpha} L b\right)-\frac{m_{b}^{2}}{m_{Z^{\prime}}^{2}} (\bar{s} R b)( \bar{s} R b)\right]\;, \label{mixing}
	\end{eqnarray}
where we have used $q^2 = m^2_{B_s}$.  In the limit $m_{Z'} >> m_{B_s}$, the second term in the bracket can be similarly neglected.
Then we obtain the ratio between the modified contribution $M_{12}$ and SM contribution $M^{SM}_{12}$
\begin{equation}
\frac{M_{12}}{M^{SM}_{12}}=1+\frac{\tilde{g}^{2}}{m_{Z^{\prime}}^{2}}\left(\frac{V_{L\; 22}^{d} V_{L\;3 2}^{d *}}{V_{tb} V_{t s}^{*}}\right)^{2}\left(\frac{9 g^{2} S_{0}}{16 \pi^{2} v^{2}}\right)^{-1} = 1+ {m^2_{Z'}\over \tilde g^2}\left({3 G_F
\alpha_{em} C_9\over \sqrt{2} \pi |V^l_{L\;22}|^2}\right)^2 \left(\frac{9 g^{2} S_{0}}{16 \pi^{2} v^{2}}\right)^{-1}\;,
\end{equation}
where  $S_{0}$  is Inami-Lim function in the SM with value $\simeq 2.3$~\cite{Lenz:2010gu,Buras:2012jb}. The mixing amplitude
$M_{12}$ is related to the mass difference by $\Delta m_{B_{s}}=2\left|M_{12}\right|$.
Using the experimental data $\Delta m_{B_{s}}=17.741\pm 0.020\;\mbox{ps}^{-1}$ in Ref.~\cite{Zyla:2020zbs} and SM prediction $\Delta m_{B_{s}}^{SM}=18.5^{+1.2}_{-1.5}\;\mbox{ps}^{-1}$ in Ref.~\cite{King:2019lal}, we obtain $ M_{12}/ M^{SM}_{12}=0.959_{-0.078}^{+0.062}$. Combining the constraints from $C_9$ operator in Eq.~(\ref{C9constrains}), we can obtain
\begin{equation}
\frac{\tilde{g}}{m_{Z^{\prime}}} =
{4vG_F \alpha_{em} \over g} {-C_9 \over  |V^l_{L\; 22}|^2}
\frac{1}{\sqrt{2S_{0}\left(\frac{M_{12}}{M^{SM}_{12}}-1\right) }}\;,\;\;\;\;\;\;\;
\left|\frac{V_{L\; 22}^{d} V_{L\;32}^{d *}}{V_{t b} V_{t s}^{*}}\right| =
\frac{3\sqrt{2}g^2 S_{0}}{16\pi v^2 G_F \alpha_{em}}  \left(\frac{M_{12}}{M^{SM}_{12}}-1\right)
\frac{ |V^l_{L\; 22}|^2} {-C_9}\;.
\end{equation}
Then we can obtain at $2 \sigma$ level the lower bound  for $\tilde{g}/m_{Z^{\prime}}\ge 1.08\times 10^{-4}$  and the upper bound for  $\left|(V_{L\; 22}^{d} V_{L\;32}^{d *})/(V_{t b} V_{t s}^{*})\right| \le 2.53$ with
$V^l_{L\; 22} = 1$.
The excluded parameter space in the $\tilde{g}-m_{Z'}$ plane is shown in yellow in Fig.~\ref{C9VL22100}. The region between two green dashed lines is the allowed space to
explain the $b \to s \mu \bar \mu$ anomalies. The up line corresponds to $\cos \theta_q \simeq 10^{-5}$( $\tilde{g}/m_{Z^{\prime}} \simeq 0.015\mbox{GeV}^{-1}$) and  the lower one is for $\cos \theta_q \simeq 0.4$ ( $\tilde{g}/m_{Z^{\prime}} \simeq 7 \times 10^{-5}\mbox{GeV}^{-1}$).

Combining all these constraints, we find that there exist  suitable regions to  explain the $b \to s \mu \bar \mu$ anomalies and satisfy all the constraints simultaneously. Therefore, $U(1)_{B_2-L_\mu}$ model can provide solutions to the $B$ anomalies  in large $m_{Z'}$ case.

Now we give some comments on the small $m_{Z^{\prime}}$ region.
For $D-\bar D$ mixing in MeV scale, the enhancement factor $m^2_{c}/m^2_{Z^{\prime}}$ in Eq.~(\ref{DDbarmixing})
makes the second term in the bracket dominant. This modifies the mass difference as
\begin{eqnarray}
\Delta m_D=-\frac{5}{8}\frac{m_c^2}{m_c^2-m_{Z'}^2}\frac{m^2_{D}}{(m_u+m_c)^2}{2\over 3}f_D^2 B_D m_D c(m_{Z^\prime}){\tilde g^2 \over 18m^2_{Z^\prime}} \left[(V_{us} \cos \theta_q-V_{ub}\sin \theta_q)(V^*_{cs} \cos \theta_q-V^*_{cb}\sin \theta_q)\right]^2\;.
\end{eqnarray}
Comparing to Eq.~(\ref{DD}), it only multiplies an additional factor $-1.35$. The extra negative sign can be compensated by corresponding CKM matrix element. Numerically for $m_{Z^{\prime}} = 0.1$ GeV and $\tilde{g} = 10^{-3}$, we can obtain $\Delta m_D= 1.21 \times 10^{10} s^{-1}$ which is consistent with the experimental data within $2\sigma$ error.

Similarly, for $B_s-\bar B_s$ mixing in  MeV scale, the amplitude is modified due to the enhancement factor $m^2_{b}/m^2_{Z^{\prime}}$ in Eq.~(\ref{mixing}) as
\begin{eqnarray}
	\label{M12C9}
	\frac{M_{12}}{M^{SM}_{12}} &&=1-\frac{5}{8} \frac{\tilde{g}^{2}}{m_{Z^{\prime}}^{2}}\left(\frac{V_{L\; 22}^{d} V_{L\;3 2}^{d *}}{V_{tb} V_{t s}^{*}}\right)^{2}\left(\frac{9 g^{2} S_{0}}{16 \pi^{2} v^{2}}\right)^{-1} \frac{m^2_{B_{s}}}{(m_b+m_s)^2}\nonumber\\
	&&= 1-\frac{\bar m^4}{m^4_{Z^{\prime}}} {m^2_{Z'}\over \tilde g^2} {45\over 8} \left({G_F \alpha_{em} C_9 \over \sqrt{2} \pi}\right)^2  \left(\frac{9 g^{2} S_{0}}{16 \pi^{2} v^{2}}\right)^{-1} \frac{m^2_{B_{s}}}{(m_b+m_s)^2}\;.
\end{eqnarray}
In the above we have used an approximation for $C^{new}_9$ with small $Z'$ mass by taking the factor $1/(q^2-m^2_{Z'})$ in Eq.~(\ref{c9exp}) with a central value $\bar m^2 \approx 3 \mbox{GeV}^2$ for $q^2$ in relevant region $1/(\bar m^2- m^2_{Z'})$.
We find that for $m_{Z^{\prime}} = 0.1$ GeV and $\tilde{g} = 10^{-3}$, it leads to  $M_{12}/M^{SM}_{12} = -1.05$ which contradicts with the experimental data.   We have searched most parameter spaces and find that there is no solution for the $B$ anomalies satisfying $B_s -\bar B_s$ mixing constraint.

The kinematically allowed two-body decay $D \to \pi Z'$, courtesy of small $m_{Z'}$ case with MeV scale, results in $D \to \pi + E_{miss}$ signature constraining our model parameters severely. Unfortunately,  there is no dedicated experimental search for this signature yet. If one adopts the decay bound  for a massless invisible pseudoscalar $a$ with $Br(D \to \pi a)< 8\times 10^{-6}$ \cite{Zyla:2020zbs} and assumes the constraint also applies to hundred MeV $Z'$ as suggested in  Ref.~\cite{Greljo:2021npi}. This also rules out a small $Z^\prime$ mass of 300 MeV as the solution for $B$ anomalies~\cite{Greljo:2021npi}.

We conclude that the $U(1)_{B_2-L_\mu}$ model discussed here is able to solve the $B$ anomalies problem in the case of large $m_{Z'}$ with hundred GeV scale.

\section{Difficulties to simultaneously solve $(g-2)_\mu$ and $B$ anomalies in $U(1)_{B_q-L_\mu}$ models}

We have carried out detailed analysis for $U(1)_{B_2-L_\mu}$ model to solve the $(g-2)_\mu$ and $B$ anomalies separately in Sec. III and Sec. IV, respectively. To simultaneously solve the $(g-2)_\mu$ and $B$ anomalies, we just need to find out  the parameter spaces for solving the $(g-2)_\mu$ and $B$ anomalies separately  and to see if there are common regions where both anomalies can be accommodated.

Due to severe constraints from $\tau \to \mu \bar \mu \mu$, $\tau \to \mu \gamma$ and $\tau \to \mu Z'$,
only $|V^l_{L\;22}|\sim 1$  is allowed as can be seen in Fig.~\ref{contrains_MZ1} whatever the large or small $m_{Z'}$ scenarios. On the premise of $|V^l_{L\;22}|\sim 1$, neutrino trident  data then force $Z'$ mass to be less than a few hundred MeV  while explaining $(g-2)_\mu$ anomaly as can be seen in Fig.~\ref{g_2Trident}.

With the above allowed parameter space for $(g-2)_\mu$ anomaly in mind, if one further requires simultaneous solve the $b\to s \mu \bar \mu$ anomalies, one just needs to concentrate on $Z'$ mass below a few hundred MeV region to see if the required $C_9$ can be found and the other constraints are not upset. We find that to produce the required $C_9$ is relatively easy. However, the same parameter space will induce $B_s-\bar B_s$ mixing which brings in devastation for such models.
To illustrate the difficulties clearly, we adopt the specific numerical analysis for $m_{Z^{\prime}} = 0.1$ GeV with $V^{l}_{L\;22} = 1$, which can explain muon $g-2$ anomaly and satisfy neutrino trident process simultaneously. Then we find  the corresponding $\tilde{g}\simeq 10^{-3}$ under this case  satisfying the $B$ anomalies  from Fig.~\ref{g_2Trident}.  Further, with the help of Eq.~(\ref{M12C9}), we obtain ${M_{12}}/{M^{SM}_{12}} = -1.05$, which is far beyond the allowed  experimental bounds. This leads to the incompatible contradictory between explaining the $B$ anomalies and satisfying $B_s-\bar B_s$ mixing due to the large enhancement factor $m^2_{b}/m^2_{Z^{\prime}}$ for $Z'$ in MeV scale.
Therefore, we conclude that it is impossible for $U(1)_{B_2-L_\mu}$ model to solve the $(g-2)_\mu$ and $b\to s \mu \bar \mu$ induced anomalies  simultaneously.

There are several variations of $U(1)_{B_q-L_l}$ model with different combinations of $q, l=2, 3$.
The gauged $U(1)_{B_3-L_\mu}$ model is among the variants.
Similar as $U(1)_{B_2-L_\mu}$,  we can construct the following favorite structure
\begin{eqnarray}
L_{int-sb, \mu,\tau} = &&- \tilde g \left (( V^l_{L\;22} V^{l*}_{L\;22}) \bar \mu \gamma^\mu \mu
+ ( V^l_{L\;32} V^{l*}_{L\;32}) \bar \tau \gamma^\mu \tau + ( V^l_{L\;22} V^{l*}_{L\;32}) \bar \mu \gamma^\mu \tau
+ ( V^l_{L\;32} V^{l*}_{L\;22}) \bar \tau \gamma^\mu \mu \right )Z'_\mu \nonumber\\
&&+ {1\over 3} \tilde g \left (V^d_{L\; 23} V^{d*}_{L\;33} \bar s_L \gamma^\mu b_L
\right ) Z'_\mu\;. \label{zprime-B3L2}
\end{eqnarray}
Comparing to Eq.~(\ref{zprime-current}), we find that the only difference is from the quark coupling. We can conduct the similar analysis just by changing the mixing parameter $V^d_{L\; 22} V^{d*}_{L\;32}$ into $V^d_{L\; 23} V^{d*}_{L\;33}$.
Therefore, we can draw a same conclusion that $U(1)_{B_3-L_\mu}$ model can not accommodate $(g-2)_\mu$ and $B$ anomalies simultaneously.

Now we comment on another possibility modifying the lepton coupling of $U(1)_{B_q-L_l}$ model,
such as $U(1)_{B_2-L_\tau}$ model. The relevant Lagrangian can be written as
\begin{eqnarray}
L_{int-sb, \mu,\tau} = &&- \tilde g \left (( V^l_{L\;23} V^{l*}_{L\;23}) \bar \mu \gamma^\mu \mu
+ ( V^l_{L\;33} V^{l*}_{L\;33}) \bar \tau \gamma^\mu \tau + ( V^l_{L\;23} V^{l*}_{L\;33}) \bar \mu \gamma^\mu \tau
+ ( V^l_{L\;33} V^{l*}_{L\;23}) \bar \tau \gamma^\mu \mu \right )Z'_\mu \nonumber\\
&&+ {1\over 3} \tilde g \left (V^d_{L\; 22} V^{d*}_{L\;32} \bar s_L \gamma^\mu b_L
\right ) Z'_\mu\;. \label{zprime-B2L3}
\end{eqnarray}
Similarly, we can analyze this case only by  modifying $V^l_{L\;j2} V^{l*}_{L\;i2}$ into $V^l_{L\;j3} V^{l*}_{L\;i3}$.  $\tau- 3\mu$, $\tau \to \mu \gamma$  and $\tau\to \mu Z'$  force the ``j" in $V^l_{L\; j3}$ to be $j=3$. In this case there is no solution for $(g-2)_\mu$.
Therefore we conclude that $U(1)_{B_q-L_\tau}$ model can hardly explain the muon $(g-2)_\mu$ and $B$ anomalies simultaneously.

\section{Kinetic mixing effects}

So far we have neglected a possible renormalizable  kinetic mixing term between $U(1)_Y$ and $U(1)_{B_2-L_\mu}$, $(1/2)\delta B^{\mu\nu} Z'_{\mu\nu}$ with $|\delta| <1$ where $B$ is the $U(1)_Y$ gauge field.
In terms of photon field $A$ and $Z$ boson field, $B = \cos\theta_W A - \sin\theta_W Z$.  When writing into gauge fields in canonical form of the physical mass eigen-state gauge bosons, the photon $A^m$, the $Z^m$ boson  and the $Z^{\prime m}$ boson, the interaction with SM current will be modified to
\begin{eqnarray}
L_{int} &=& J^\mu_{em} A^m_\mu+ \left (-{s_\xi \delta \cos\theta_W\over \sqrt{1-\delta^2}} J^\mu_{em} +(c_\xi + {s_\xi \delta \sin\theta_W \over \sqrt{1-\delta^2}}) J^\mu_Z
+ {s_\xi \over \sqrt{1-\delta^2}}J^\mu_{Z'}\right)Z^m_\mu \nonumber\\
&+& \left (- {c_\xi \delta \cos\theta_W \over \sqrt{1-\delta^2}}J^\mu_{em} - (s_\xi -{c_\xi \delta \sin\theta_W \over \sqrt{1-\delta^2}})J^\mu_Z   + {c_\xi \over \sqrt{1-\delta^2}}J^\mu_{Z'} \right ) Z^{\prime m}_\mu\;,\label{Kinetic}
\end{eqnarray}
where
$J^\mu_{em} = - e Q^f \bar f \gamma^\mu f$, $J^\mu_Z = \bar f \gamma^\mu (g^f_V - g^f_A \gamma_5) f$ with
$g_V^f = -(g/2\cos\theta_W) (I_3^f - 2 Q^f \sin^2\theta_W)$ and $g_A^f = -(g/2\cos\theta_W) I_3^f$, and $J_{Z^{\prime }}^\mu $ is defined by Eq.~(\ref{zprime-current}).
Here $Q^f$ is the electric charge of fermion $f$ in unit $e$,
and $I^f_3$ are $1/2$, $-1/2$  and $0$ for the up and down components of $SU(2)_L$ doublet  and singlet fermions. We will drop the superscript $m$ in the boson fields in our later discussions.
And $c_\xi (s_\xi )= \cos\xi (\sin\xi)$ is with  the mixing angle $\xi$  of $Z$ and $Z'$  as~\cite{gangli}
\begin{eqnarray}
\tan 2\xi = {2 m^2_Z \delta \sin\theta_W/\sqrt{1-\delta^2} \over m^2_Z - (m^2_Z \delta^2 \sin^2\theta_W + m^2_{Z'})/(1-\delta^2)}\;.\label{xi}
\end{eqnarray}
Note that there is a resonant for $\xi$ when $m_{Z'}$ is near $m_Z$. To avoid this situation, we choose
large $Z'$ mass above 100 GeV. In this case, experimental constraints on $\delta$ and $s_\xi$ are weak with $|\delta|(|s_\xi|)<1$.

The above kinetic mixing will make a correction for muon $g-2$. We find that when ignoring the photon contribution, the largest correction  only comes from the second order $\delta^2$ with
\begin{eqnarray}
\frac{\tilde{g}^{2}}{4 \pi^{2}} \frac{m_{\mu}^{2}}{m_{Z}^{2}} \left ({m^2_Z\over m^2_Z - m^2_{Z'}} \right )^2 \delta^2\sin^2\theta_W \left(V^l_{L\; 22} V_{L\; 22}^{l*}\right)\left[V^l_{L\; 32} V_{L\;32}^{l*}\left(\frac{m_{\tau}}{m_{\mu}}-\frac{2}{3}\right)+\frac{1}{3} V^l_{L\; 22} V_{L\; 22}^{l*}\right]\;.
\end{eqnarray}
As long as $m_{Z'}>m_Z$ and $\delta<0.3$ (although we do not think it can be this large), the very limited effects can be neglected. This effect, however, can be probed by high energy colliders~\cite{dark1,dark2}.

It has been pointed out that kinetic mixing can in principle produce a $C_9^{loop}$ \cite{2007mixing} by first inducing  the one loop SM contribution to $\bar s b Z$ coupling and then  mixing $Z$ with $Z'$  to couple to $\mu\bar \mu$. The contribution can be written as
\begin{eqnarray}
\label{C9mixingcont}
	C^{loop}_9 = 2\cos\theta_W { \tilde g (V^{l}_{L\;22}V^{l*}_{L\;22})  \over e\sin\theta_W} {m^2_{Z} \over m^2_{Z'}}{c_\xi \over \sqrt{1-\delta^2}}\left(s_\xi-c_\xi {\delta \sin\theta_W \over \sqrt{1-\delta^2}}\right) C_0(m^2_t/m_W^2)\;,
\end{eqnarray}
where
\begin{eqnarray}
C_0(x) = {x\over 8} \left ({6-x\over (1- x)} +{3x + 2\over (1-x)^2} \ln x \right )\;.
\end{eqnarray}
We have carried out a unitary calculation for the loop contribution using results from Ref.~\cite{h-g-t}. Our expression for $C^{loop}_9$ differs from that obtained in Ref.~\cite{2007mixing}.

Numerically, we find that the produced $C_9$  is too small to generate the required value to solve $b \to s \mu \bar \mu$ induced anomalies. Thus, we try to make $s_\xi$ substantially away from 0 to obtain a sizable contribution for  $C_9$. However, in this case the term $J^\mu_{Z'}Z^m_\mu$ in Eq.~(\ref{Kinetic}) will be important to affect  $Z$ interactions with SM particles, which is infeasible because $Z$ has been severely constrained by precision test data to very close to SM predicted interactions.

We conclude that kinetic mixing will not be able to help much to deal with the challenges to solve $(g-2)_\mu$ and $B$ anomalies simultaneously.

\section{ Discussions and Conclusions}

The muon $(g-2)_\mu$ and $b\to s \mu \bar \mu$ induced anomalies, both belonging to new physics beyond SM, attract much of attention.  These tantalizing anomalies share the same feature involving the second generation of charged lepton, which indicates that they can be correlated by new interaction specifically related to muon. We study the possibility of using
gauged flavor specific $U(1)_{B_q-L_\mu}$ model to explain the $(g-2)_\mu$ and $B$ anomalies. We find although for $U(1)_{B_q-L_\mu}$ models there is still parameter space to provide solutions for separately explaining the $(g-2)_\mu$ and $B$ anomalies, there exists no parameter region for such models to solve both the anomalies simultaneously, after taking into account existing constraints from $\tau \to \mu \gamma$, $\tau \to 3 \mu$, $\tau \to \mu Z' $, neutrino trident  and $B_s - \bar B_s$ data.

We started with $U(1)_{B_2-L_\mu}$ model to illustrate the concrete details about solving muon $(g-2)_\mu$ and $B$ anomalies separately and simultaneously.
On the one hand, to satisfy severe constraints from other processes, such as  $\tau \to \mu \gamma$, $\tau \to 3 \mu$, $\tau \to \mu Z' $ and  neutrino trident, the only suitable solution to $(g-2)_\mu$ anomaly is $V^l_{L\;22} = 1$ and $m_{Z'}<300$ MeV.
On the other hand, the above viable small $m_{Z'}$ scale will be ruled out by the constraints from $B_s$ mixing and $D\to \pi Z'$. The only existing  suitable region to explain the $B$ anomalies separately is in large $m_{Z'}$ scale.
Unfortunately, we found that there exists no common region to accommodate these two anomalies. Therefore,  we conclude that the $U(1)_{B_2-L_\mu}$ model can not explain $(g-2)_\mu$ and $B$ anomalies simultaneously.

We also found that variations of  $U(1)_{B_q-L_l}$ model are impossible to realize the above accommodating purpose.
For $U(1)_{B_3 - L_\mu}$ model, it only modifies  the quark coupling so that the difficulties will appear again.
For $U(1)_{B_q - L_\tau}$ model, it will not provide the solution for $(g-2)_\mu$ when considering other constraints from $\tau- 3\mu$, $\tau \to \mu \gamma$  and $\tau\to \mu Z'$, which force $V^l_{L\; 33}=1$.
Therefore, variations of $U(1)_{B_q-L_l}$ model can also not explain $(g-2)_\mu$ and $B$ anomalies simultaneously. We also studied kinetic mixing effects, but found that neither can it  help to solve the problems.

\section*{Acknowledgments}
This work was supported in part by Key Laboratory for Particle Physics, Astrophysics and Cosmology, Ministry of Education, and Shanghai Key Laboratory for Particle Physics and Cosmology (Grant No. 15DZ2272100), and in part by the NSFC (Grant Nos. 11735010, 11975149, and 12090064). XGH was supported in part by the MOST (Grant No. MOST 106-2112-M-002-003-MY3 ).

\end{document}